\documentclass[namedreferences]{solarphysics}
\usepackage[optionalrh]{spr-sola-addons} % For Solar Physics 
\usepackage{graphicx}        % For eps figures, newer & more powerfull
\usepackage{subfig}
\usepackage{amssymb}        % useful mathematical symbols
\usepackage{color}           % For color text: \color command
\usepackage{url}             % For breaking URLs easily trough lines
\def\ion#1#2{#1$\;${\tiny\rm{#2}}\relax} % define the roman ionization symbol

\newcommand{\arcsec}{$^{\prime\prime}$}

\newcommand{\FeXII}{\ion{Fe}{XII}}
\newcommand{\FeXIII}{\ion{Fe}{XIII}}

\newcommand{\kms}{km s$^{-1}$}

\begin{document}

\begin{article}

\begin{opening}

\title{Oscillations in active region fan loops: Observations from EIS/{\it Hinode} and AIA/SDO}

\author{S.~\surname{Krishna Prasad}$^{1}$\sep
        D.~\surname{Banerjee}$^{1}$\sep
        Jagdev~\surname{Singh}$^{1}$      
       }
\runningauthor{Prasad, Banerjee, Singh}
\runningtitle{Oscillations in fan loops}

   \institute{$^{1}$ Indian Institute of Astrophysics, Bangalore-560034, India.\\
                     email: \url{dipu@iiap.res.in} }

\begin{abstract}
Active region fan loops in AR 11076 were studied, in search of oscillations, using high cadence spectroscopic observations from \textit{Extreme-ultraviolet Imaging Spectrometer} (EIS) on board {\it Hinode} combined with imaging sequences from the \textit{Atmospheric Imaging Assembly} (AIA) on board SDO. Spectra from EIS were analyzed in two spectral windows, \FeXII\ 195.12~\AA\ and \FeXIII\ 202.04~\AA\ along with the images from AIA in 171~\AA\ and 193~\AA\ channels. We find short ($<$3~min) and long ($\approx$9~min) periods at two different locations. Shorter periods show oscillations in all the three line parameters and the longer ones only in intensity and Doppler shift but not in line width. Line profiles at both these locations do not show any visible blue-shifted component and can be fitted well with a single Gaussian function along with a polynomial background. Results using co-spatial and co-temporal data from AIA/SDO do not show any significant peak corresponding to shorter periods, but longer periods are clearly observed in both 171~\AA\ and 193~\AA\ channels. Space-time analysis in these fan loops using images from AIA/SDO show alternate slanted ridges of positive slope, indicative of outward propagating disturbances. The apparent propagation speeds were estimated to be 83.5 $\pm$ 1.8~\kms\ and 100.5 $\pm$ 4.2~\kms, respectively, in the 171~\AA\ and 193~\AA\ channels. Observed short period oscillations are suggested to be caused by the simultaneous presence of more than one MHD mode whereas the long periods are suggested as signatures of slow magneto-acoustic waves. In case of shorter periods, the amplitude of oscillation is found to be higher in EIS lines with relatively higher temperature of formation. Longer periods, when observed from AIA, show a decrease of amplitude in hotter AIA channels which might indicate damping due to thermal conduction owing to their acoustic nature.
\end{abstract}
\end{opening}
%-------------------------------------------------
\section{Introduction}
Detection of waves and oscillations in the outer solar atmosphere helps us not only to understand their role in coronal heating but also to remotely diagnose the properties of the corona - coronal seismology (\opencite{1970PASJ...22..341U}; \opencite{1984ApJ...279..857R}). Active region oscillations had been reported by several authors, both from imaging (\opencite{1999SoPh..190..249N}; \opencite{2000A&A...355L..23D}; \opencite{2001A&A...370..591R}; \opencite{2003A&A...404L...1K}; \opencite{2006A&A...448..763M}) and spectroscopic (\opencite{1999A&A...347..355I}; \opencite{2001A&A...368.1095O}, \opencite{2002A&A...387..642O}, \opencite{2009A&A...494..355O}; \opencite{2008ApJ...681L..41M}, \opencite{2010ApJ...713..573M}; \opencite{2009ApJ...696.1448W}, \citeyear{2009A&A...503L..25W}) observations. \inlinecite{1999A&A...347..355I} and \inlinecite{2001A&A...368.1095O} find oscillations with periods as short as 1~min using observations from CDS/SOHO. \inlinecite{2002A&A...387..642O} observed oscillations over a sunspot region at different heights in the atmosphere with frequencies ranging from 5.4~mHz (185~s) to 8.9~mHz (112~s) using both CDS/SOHO and TRACE. \inlinecite{2009A&A...494..355O} reported active region oscillations over a range of frequencies from 2~mHz (500~s) to 154~mHz (6.5~s) using the \textit{Extreme-ultraviolet Imaging Spectrometer} (EIS) on board {\it Hinode}. They also reported that the frequencies higher than 8~mHz are observed preferentially at the boundaries of bright plage regions. All of them interpreted these oscillations as either slow magneto-acoustic waves or fast mode waves.
\paragraph*{}
Extended loop regions in the active region, were commonly found to host propagating disturbances, which enhanced the interest in studying these regions. Active region fan loops were widely studied using imaging observations, mainly from TRACE ({\it e.g.} \opencite{1999SoPh..187..261S}; \opencite{2003A&A...404L...1K}; \opencite{2000A&A...355L..23D}, \opencite{2002SoPh..209...61D}, \citeyear{2002A&A...387L..13D}) and EIT/SOHO \cite{1999SoPh..186..207B}. Average values of the propagation speed, amplitude relative to the background and periodicities of these disturbances are 99.7$\pm$3.9~\kms, 3.7$\pm$0.2\% and 284$\pm$10.4~s, respectively (see the review by \opencite{2009SSRv..149...65D}). Simultaneous observations from TRACE and CDS/SOHO by \inlinecite{2003A&A...404L..37M}, also reveal intensity oscillations of 5~min periodicity. \inlinecite{2009A&A...503L..25W} reported 12~min and 25~min periods in these loops, both in intensity and velocity, using data from EIS/{\it Hinode}. All these observed propagating disturbances are interpreted in terms of propagating slow magneto-acoustic waves which led to various applications in coronal seismology. There are also a few reports on outflows at the edges of active region which can contribute significantly to the slow solar wind (\opencite{2007Sci...318.1585S}; \opencite{2008ApJ...676L.147H}; \opencite{2009ApJ...706L..80M}). Recently \inlinecite{2010ApJ...722.1013D} pointed out that the oscillations in intensity and velocity, in some cases, are accompanied by in-phase oscillations in line width of the same period when the line profiles are fitted with a single Gaussian. They show that the presence of faint quasi-periodic upflows driven from below leading to an additional blue-shifted component in the line profile can cause oscillations in all the three line parameters. Hence, it is difficult to differentiate between waves and flows using only imaging data. \inlinecite{2011ApJ...727L..37T} also reported propagating disturbances supporting the quasi-periodic upflow scenario using the simultaneous observations from XRT and EIS. These results highlight the need for simultaneous imaging and spectroscopic observations, to characterize the propagating disturbances properly. More recently, \inlinecite{2011ApJ...737L..43N}, using data from EIS/{\it Hinode}, reported the observation of propagating slow mode waves along the continuous outflow. They see an increasing correlation between intensity and velocity disturbances as we move away from the base of the outflow region. In this article, we report observations of oscillations with short ($<$ 3~min) and long periods ($\approx$9~min) in an active region fan loop system, using simultaneous observations from the EIS/{\it Hinode} and the \textit{Atmospheric Imaging Assembly} (AIA) on board SDO. We also discuss the possible interpretation of these oscillations.

\section{Observations}
 \label{S-Observations}
\begin{figure}
 \includegraphics[width=8.6cm,angle=90]{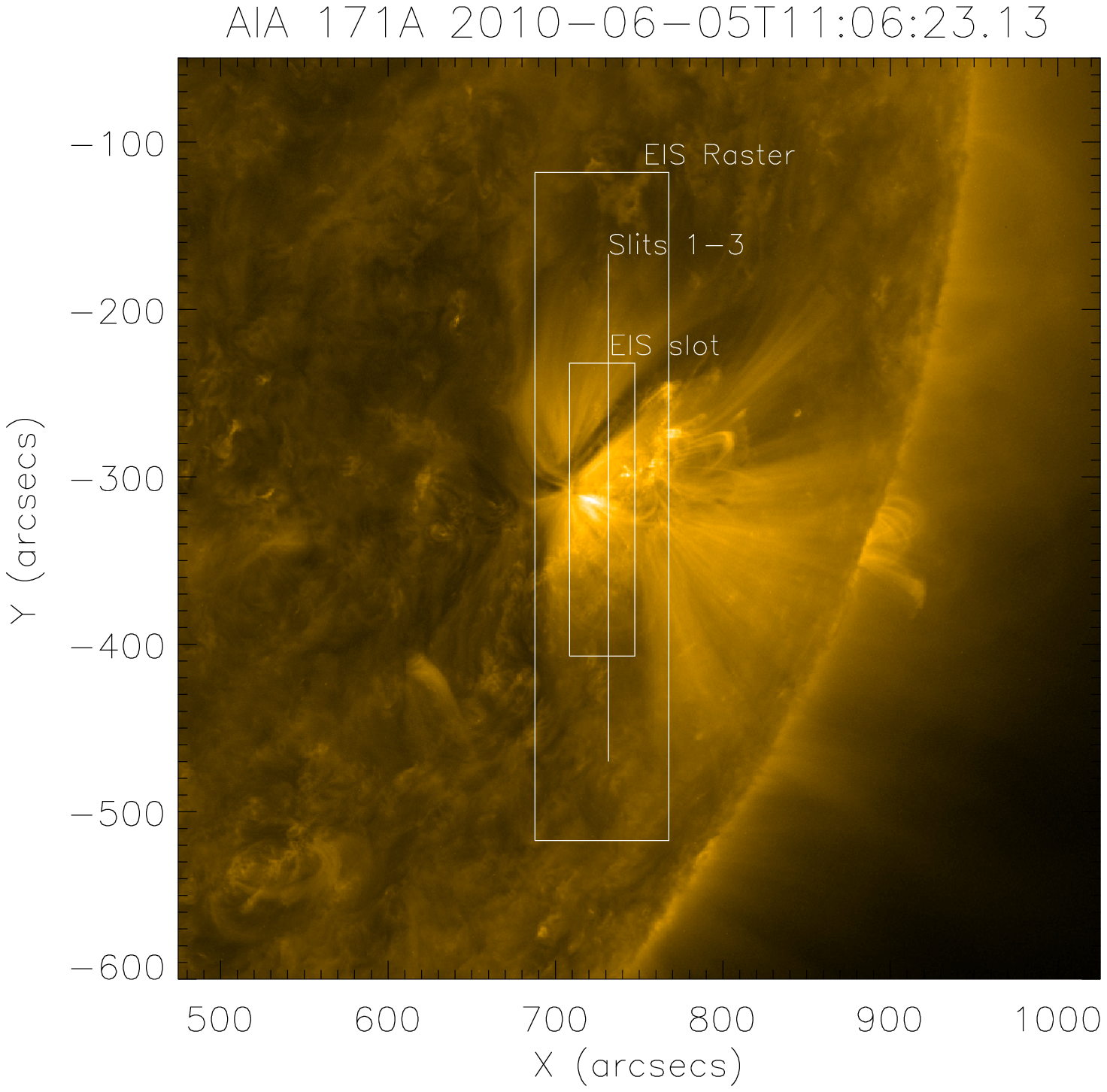}
 \includegraphics[height=3.2cm,angle=90]{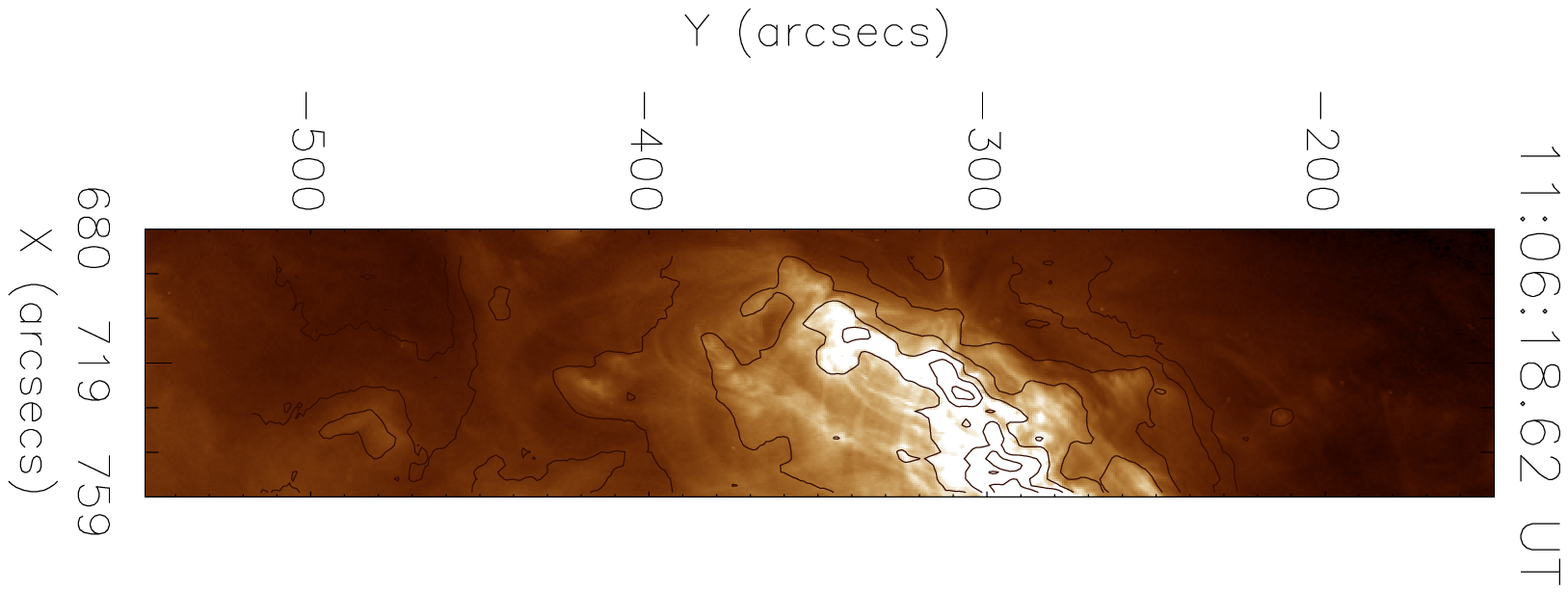}
\caption{{\it Left:} A snapshot of the region covering AR 11076 taken during the time of EIS raster using AIA/SDO in 171~\AA\ channel. Overplotted outer box displays the region covered by EIS raster, inner box indicates the location of EIS 40\arcsec\ slot and the vertical line denotes the position of EIS slit used in sit-and-stare mode. There were three sets of observations in this mode each with $\approx$10~min duration at the same location. {\it Right:} Portion of the AR covered by EIS raster, taken in the 193~\AA\ channel of AIA. Overplotted curves in black are intensity contours constructed from EIS raster after accounting for the offsets between the two instruments. This panel illustrates the goodness of the alignment.}
\label{cntxt}
\end{figure}

Spectroscopic data from the EIS (\opencite{2007SoPh..243...19C}) on board {\it Hinode} (\opencite{2007SoPh..243....3K}) combined with co-spatial and co-temporal imaging sequence from the AIA (\opencite{2012SoPh..275...17L}) on board SDO, are used in this work to study oscillations in active region fan loops. The dataset from EIS was obtained under the Hinode Operation Plan, HOP 156. The target region of EIS observations is NOAA active region (AR) 11076 when it is close to the west limb on June 5, 2010. EIS observations leading to this particular data set started with a context raster with a 2\arcsec\ slit starting at 10:57:27 UT on June 5, 2010, covering a region of 82\arcsec\ $\times$ 400\arcsec~, followed by a temporal sequence of 150 exposures with 40\arcsec\ slot at nearly 5~s cadence, which is again followed by three consecutive sets of sit-and-stare observations taken with 1\arcsec\ slit, each of 100 exposures at $\approx$6~s cadence that completed the observational sequence. The locations of the slits, slot and the region covered by the raster are marked over the subfield of an AIA image, which is shown in left panel of  Figure~\ref{cntxt}. It can be seen that, part of the slit used in sit-and-stare mode, is located over the active region fan loops. Our primary focus in this work is only on sit-and-stare observations by this slit. These observations are taken in 5 spectral lines of EIS, out of which, two lines \FeXII\ 195.12~\AA\ and \FeXIII\ 202.04~\AA\ are used in this work. The other lines were ignored either due to poor signal or due to blending issues.

\paragraph*{}
We followed the standard reduction procedure in preparing the EIS data suitable for analysis using the routines available in Solar SoftWare (SSW). This procedure involves corrections for dark current and detector bias, removal of warm, hot and saturated, and dusty pixels, removal of cosmic rays, and applying a radiometric calibration which results in intensities in absolute units. The spectra were then fitted with single Gaussian line profiles plus a background to get the spectral parameters namely, line intensity, peak position, and line width. Before fitting, we did a three-pixel binning in the spatial dimension along the slit to improve the signal. The fitting task for the whole time sequence is automatically done using \verb+eis_auto_fit.pro+ routine. This routine corrects for the EIS slit tilt and orbital variation in the line centroids. This also does the absolute wavelength calibration using the method described in \inlinecite{2010SoPh..266..209K} and reference wavelengths for the Doppler shift 
measurements were taken from rest wavelengths listed in \inlinecite{2008ApJS..176..511B}. The maximum shifts in the position of the slit during our observations, due to instrument jitter are -0.86\arcsec\ and +0.12\arcsec\ in X and -2.3\arcsec\ and +0.64\arcsec\ in Y directions, as given by the routine \verb+eis_jitter.pro+. Since the effective spatial resolution in Y-direction is 3\arcsec\ (binned over three pixels), the effects of jitter may be small. An isolated bright structure along the slit does not show any visible shifts in its position during the entire observation confirming this.

\paragraph*{}
Subfield images from AIA/SDO, covering the active region, in two coronal channels centered at 171~\AA\ and 193~\AA\, are also used. A total of 900 images ($\approx$3~hrs duration at 12~s cadence), in each channel, roughly starting at 10:00:11 UT, covering the EIS observation time, are used here. The initial data set was at level 1.0 which has already been processed to make basic corrections like removal of dark current, cosmic spikes, bad pixels etc. We then use the \verb+aia_prep.pro+ routine, available in SSW, to adjust the plate scales and roll angles in different channels to a common value. This transforms the data from level 1.0 to level 1.5. All the subfield images are tracked for solar rotation taking the reference time as the time when slit is roughly at the center of the raster.

\paragraph*{}
Pointing offsets between EIS/{\it Hinode} and AIA/SDO were calculated using the intensity cross-correlation technique to overlay the spectroheliogram constructed from EIS raster in \FeXII\ 195.12~\AA\ line on the AIA subfield, closest in time, taken through 193~\AA\ channel. Panel on the right in Figure~\ref{cntxt} shows the EIS raster in 195.12~\AA\ line overplotted on AIA subfield, after applying the offsets. Visual inspection by zooming into the smaller features tells that the alignment should be accurate at least up to a couple of arc seconds. 

\section{Analysis and Results}
\begin{figure}
 \includegraphics[width=5.6cm,angle=90]{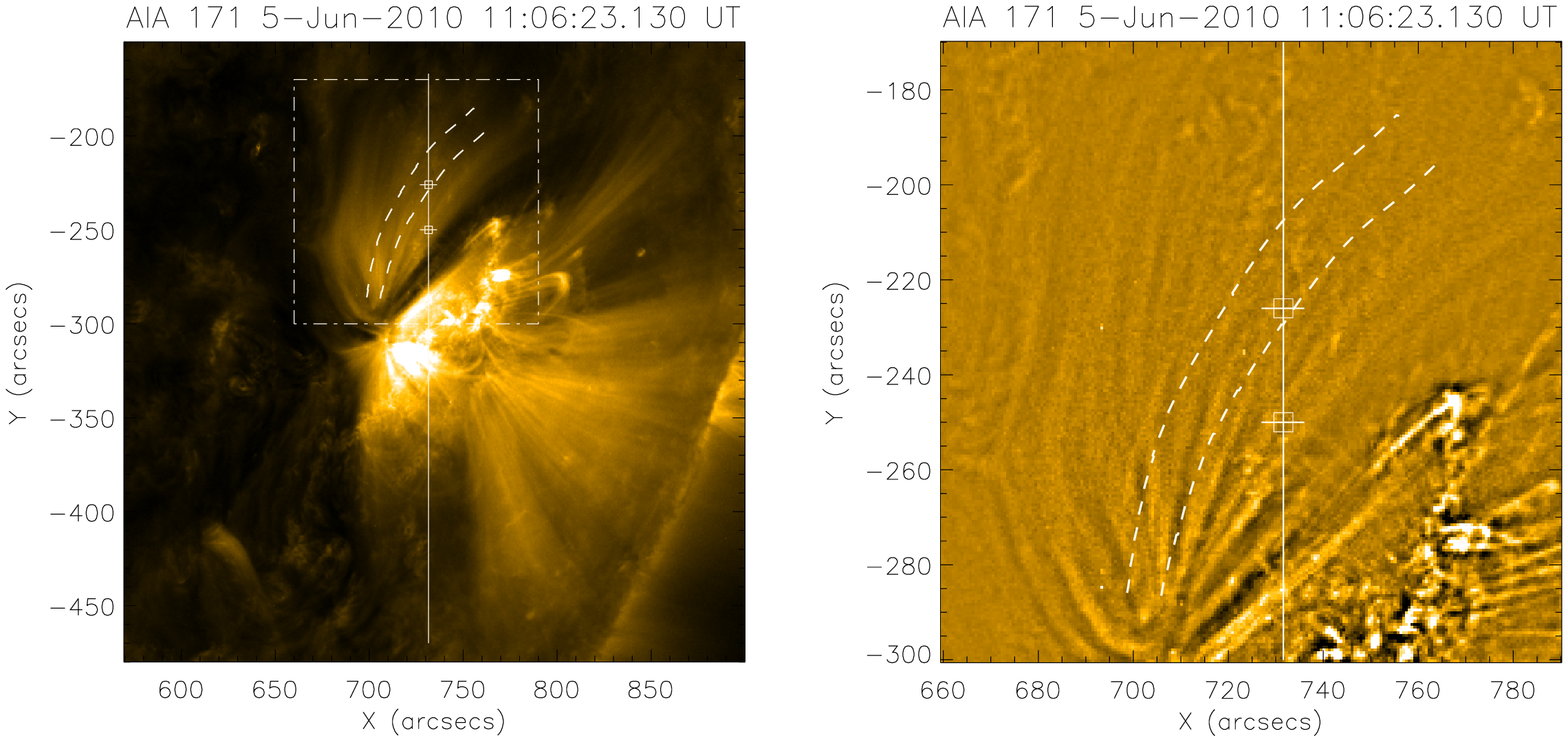}\\
 \includegraphics[width=5.6cm,angle=90]{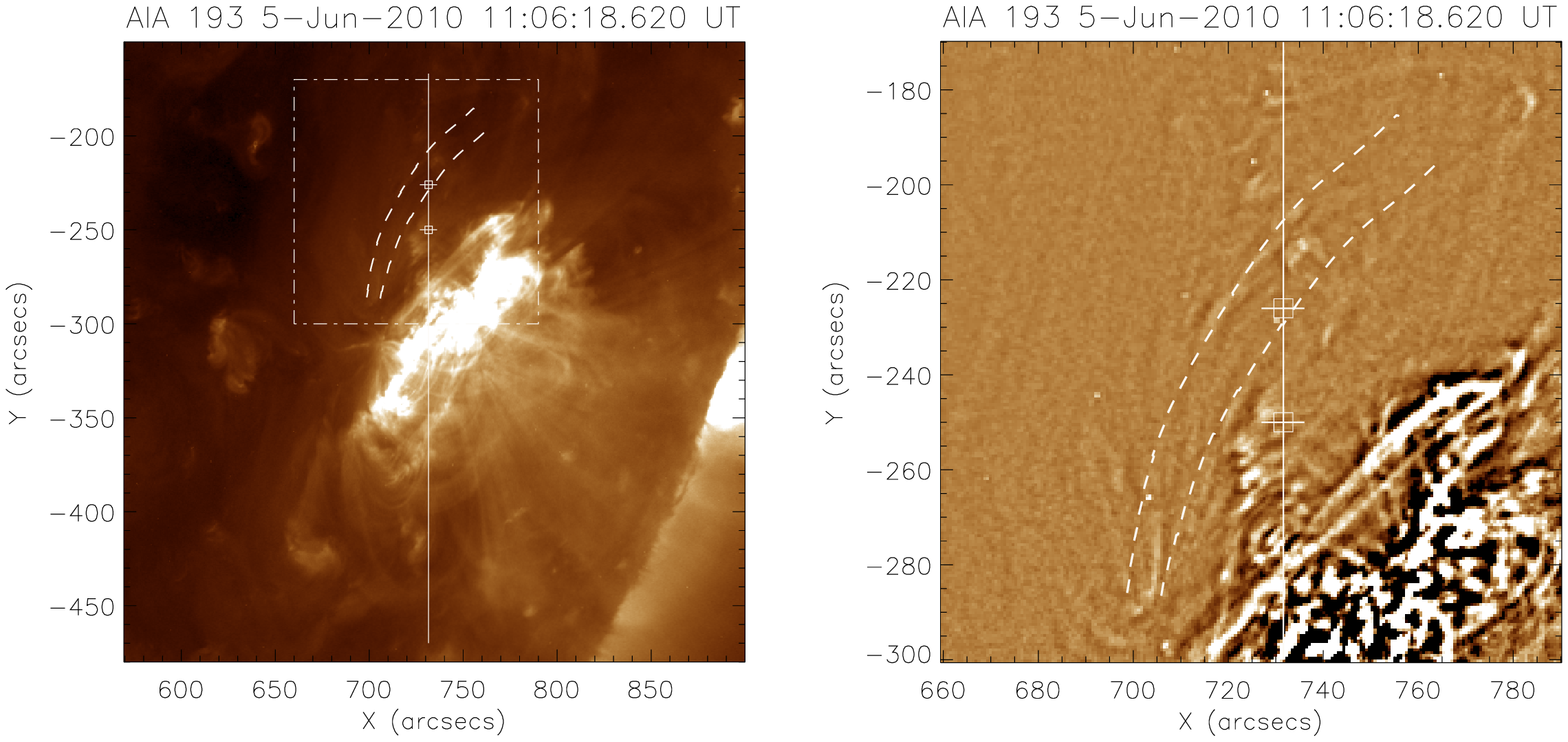}
\caption{Snapshots of the active region (left panels) taken during the raster time, in 171~\AA\ (top) and 193~\AA\ (bottom) channels of AIA. The region enclosed by the box marked with dash-dotted lines in these images, is processed using the unsharp masking technique to show the fine structure, and displayed on right for a closer view. Unsharp masking is done by taking a boxcar smooth of 8$\times$8 pixels and subtracting it from original. In each of these panels, the vertical line marks the slit position and the horizontal cuts correspond to the locations over EIS slit displaying significant oscillatory behavior. The top and bottom cut locations correspond to solar-$Y$ positions of -226\arcsec\ and -250\arcsec\ respectively. Overplotted boxes at each of these cuts, represent the corresponding region from AIA used in the analysis. The dashed curves enclose a portion of the fan loop system used in the space-time analysis.}
\label{marks}
\end{figure}

\begin{figure}
 \includegraphics[width=12cm]{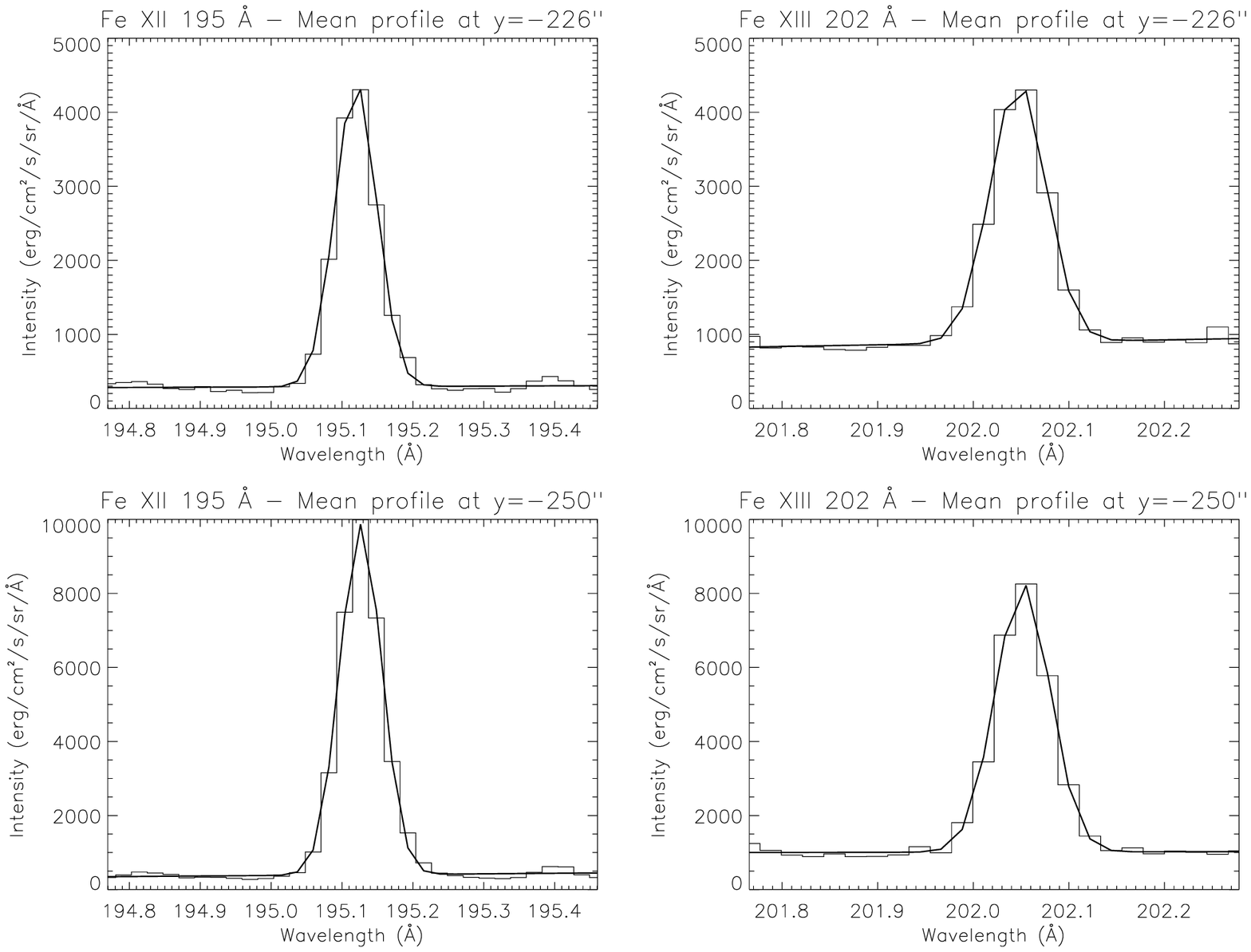}\\
\caption{Line profiles averaged over time of the sit-and-stare observation, taken in EIS spectral lines, \FeXII\ 195.12~\AA\ (left) and \FeXIII\ 202.04~\AA\ (right) using 1\arcsec\ slit. Panels in the top row correspond to the top cut location and in the bottom row correspond to the bottom cut location in Figure~\ref{marks}. In each of these panels, the histogram is the mean of the recorded line profiles and the overplotted solid line is the single Gaussian fit considering a polynomial background. This figure shows that the fits are considerably good and there is no visible extra blue-shifted component in any of these profiles.}
\label{profiles}
\end{figure}

In this section, we discuss the techniques used to detect the oscillations and their properties, both from AIA/SDO and EIS/{\it Hinode} data. Figure~\ref{marks} displays the AIA images, taken during EIS raster observation, in two channels 171~\AA\ (top) and 193~\AA\ (bottom), with the locations of our interest marked. In both rows, images on the right are produced from those on the left by subtracting an 8$\times$8 pixel boxcar smoothed image of each from itself. In all the panels, the vertical line represents the location of the EIS slit, horizontal cuts mark the position of the EIS pixels studied and the curved lines enclose the loop region analyzed from AIA data. Square boxes marked over each cut also represent the corresponding AIA region analyzed. The locations of these two cuts, bottom and top, correspond to -250\arcsec\ and -226\arcsec~, respectively, in solar-$Y$.
Mean (averaged over the time series) line profiles at these locations, in two EIS lines, are shown in Figure~\ref{profiles}. Line profiles shown in the top row, corresponding to solar-$Y \approx$ -226\arcsec~, are averaged over all the three sit-and-stare sets while those in the bottom row corresponding to solar-$Y \approx$ -250\arcsec~, are averaged over only the third set, for the reasons to be explained in Section~\ref{S-Discussion}. We find oscillations of short periods ($<$ 3~min) at the bottom cut location and long periods ($\approx$ 9~min) at the top cut location. It should be noted that the terms we use short and long are relative and used only with reference to the observed periodicities from this particular dataset.

\subsection{Short period oscillations}
 \label{S-Short period oscillations}
One of the principal aims of taking the sit-and-stare observations at a high cadence ($\approx$6~s) is to detect and understand the nature of oscillations with the periods of the order of 1~min or less. Due to telemetry restrictions, we had to limit the observations to four spectral windows and out of them two, namely, \FeXII\ 195.12~\AA\ and \FeXIII\ 202.04~\AA\, are used in the current investigation. As mentioned in the previous section, there were three sets of scans in sit-and-stare mode at the location marked by the vertical line in Figures~\ref{cntxt} and \ref{marks}. The third set of these shows significant oscillation with shorter periodicity, at a pixel location marked by the bottom cut in Figure~\ref{marks}. Light curves in intensity, velocity and line width are generated at this pixel location in the two EIS lines. A square region (see Figure~\ref{marks}) of dimension 3\arcsec\ centered around this pixel has been chosen from each AIA channel (171~\AA\ and 193~\AA) and co-temporal light curves are constructed from intensity averaged over this region. Since the AIA data is available throughout the observation time we construct AIA light curves of slightly longer duration to increase the frequency resolution. Note that increasing the duration allows the longer periods to come in whose amplitudes are normally larger and carries more power suppressing the shorter periods. So, it is optimally chosen. 

\begin{figure}
 \centering
 \includegraphics[width=9.0cm,angle=90]{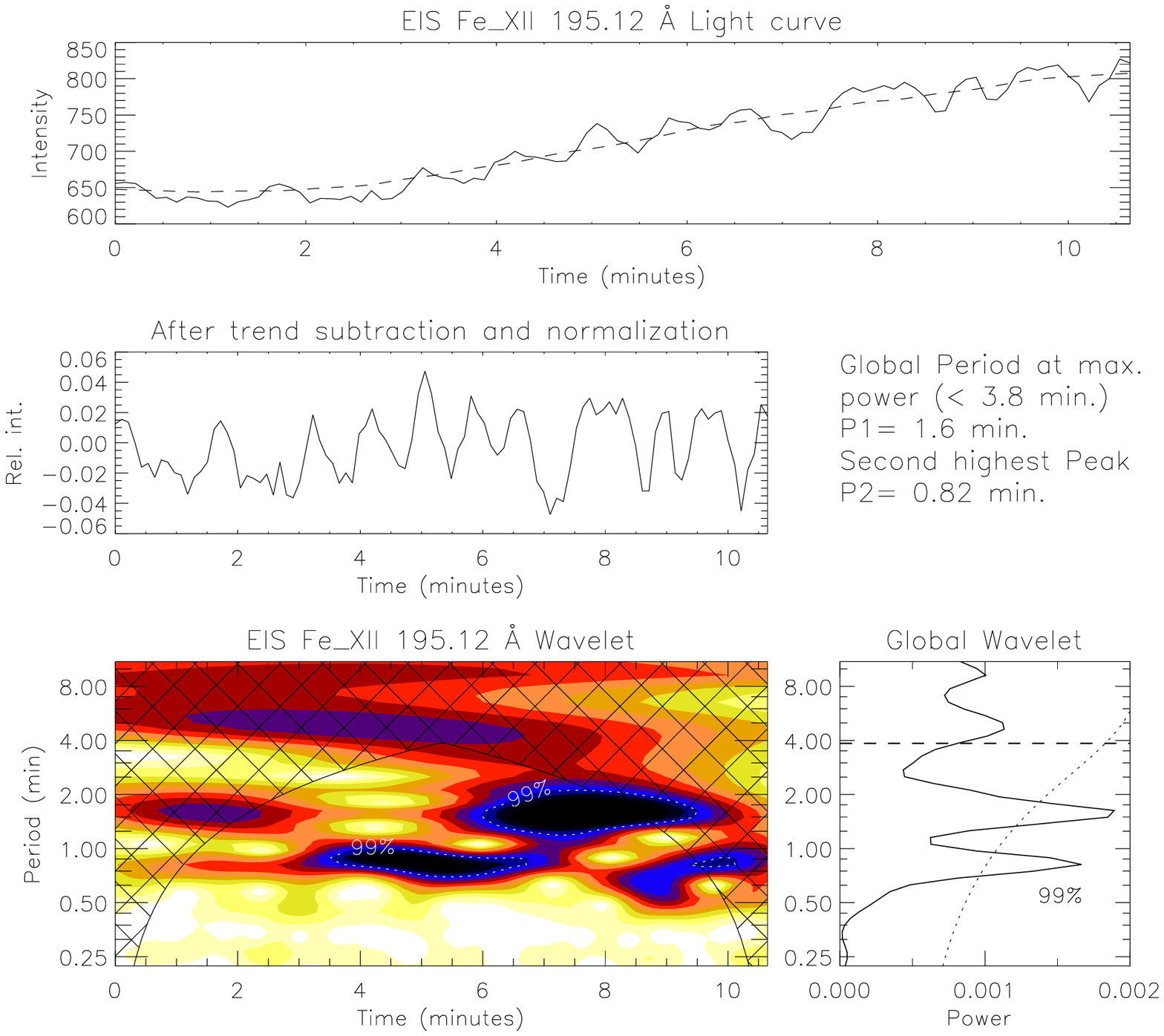}\\
 \vspace*{0.1in}
 \includegraphics[width=9.0cm,angle=90]{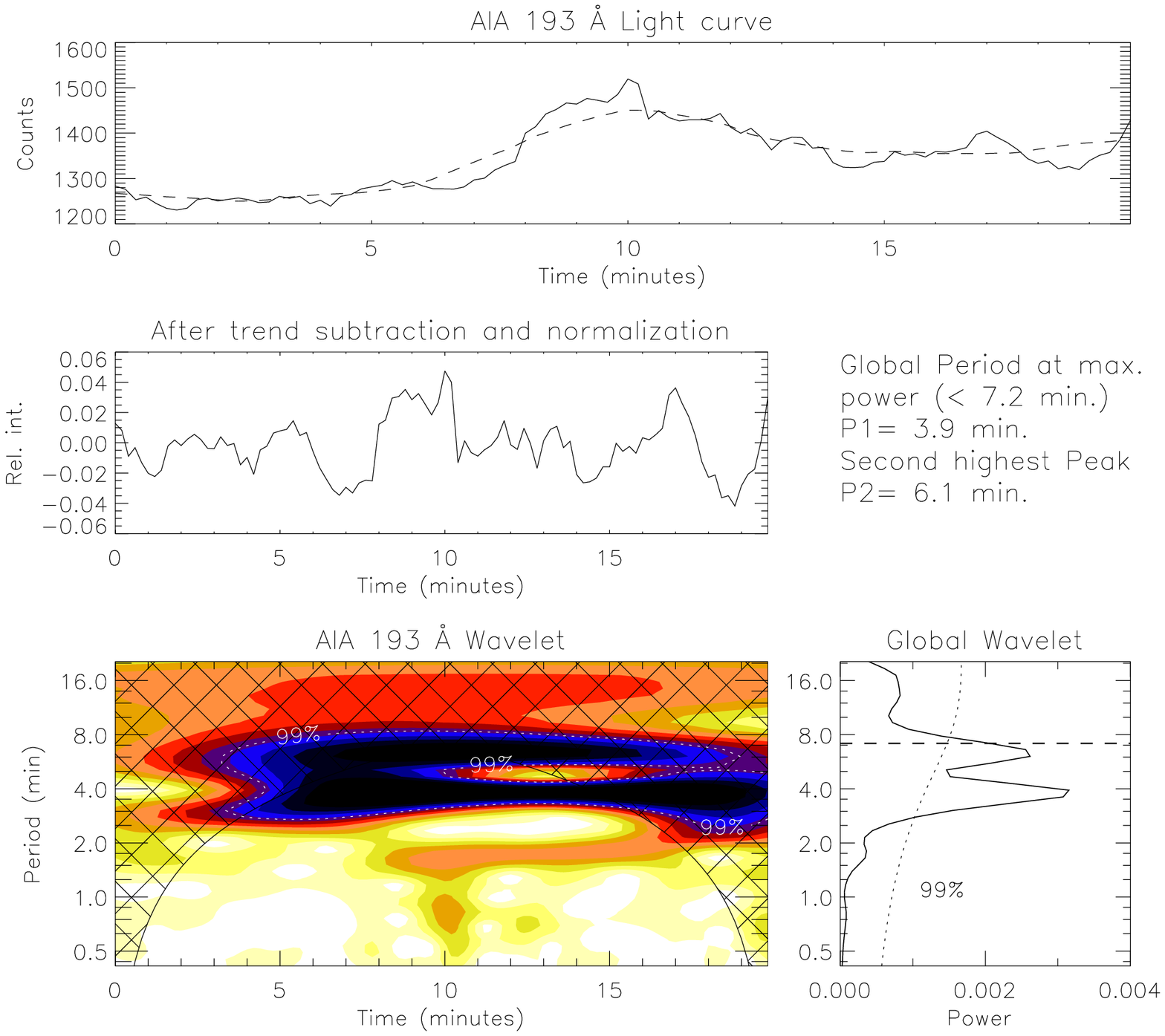}
 \caption{Results of wavelet analysis for the light curves generated from the bottom cut location (solar-$Y \approx$ -250\arcsec) in Figure~\ref{marks}. These plots are for intensity, velocity, and line width in \FeXII\ 195.12~\AA\ EIS line and intensity in 193~\AA\ channel of AIA (see the corresponding labels). In each set of plots, the top panel displays the original light curve in solid line. For EIS this is smoothed over 2 points and for AIA no smoothing is done. The dashed line overplotted on this is the $\approx$4~min running average of the original which is subtracted from it and displayed in the central panel of the set. For the intensity, the resultant is also normalized by this running average. The bottom left panel in the set shows the wavelet results which displays the different periodicities present and their temporal evolution. The color code is inverted and so black indicates the regions of strongest power. Right to this is the global wavelet plot which is the time average of the wavelet plot. The dashed horizontal line in this panel is the cutoff above which the edge effects come into play and the curved dotted line marks the 99\% significance level for a white noise process \protect\cite{1998BAMS...79...61T}. Values of the top two strongest periods are listed as P1 and P2 in the text adjacent to the middle panels.}
\end{figure}
\begin{figure}
 \ContinuedFloat
 \centering
 \includegraphics[width=9.0cm,angle=90]{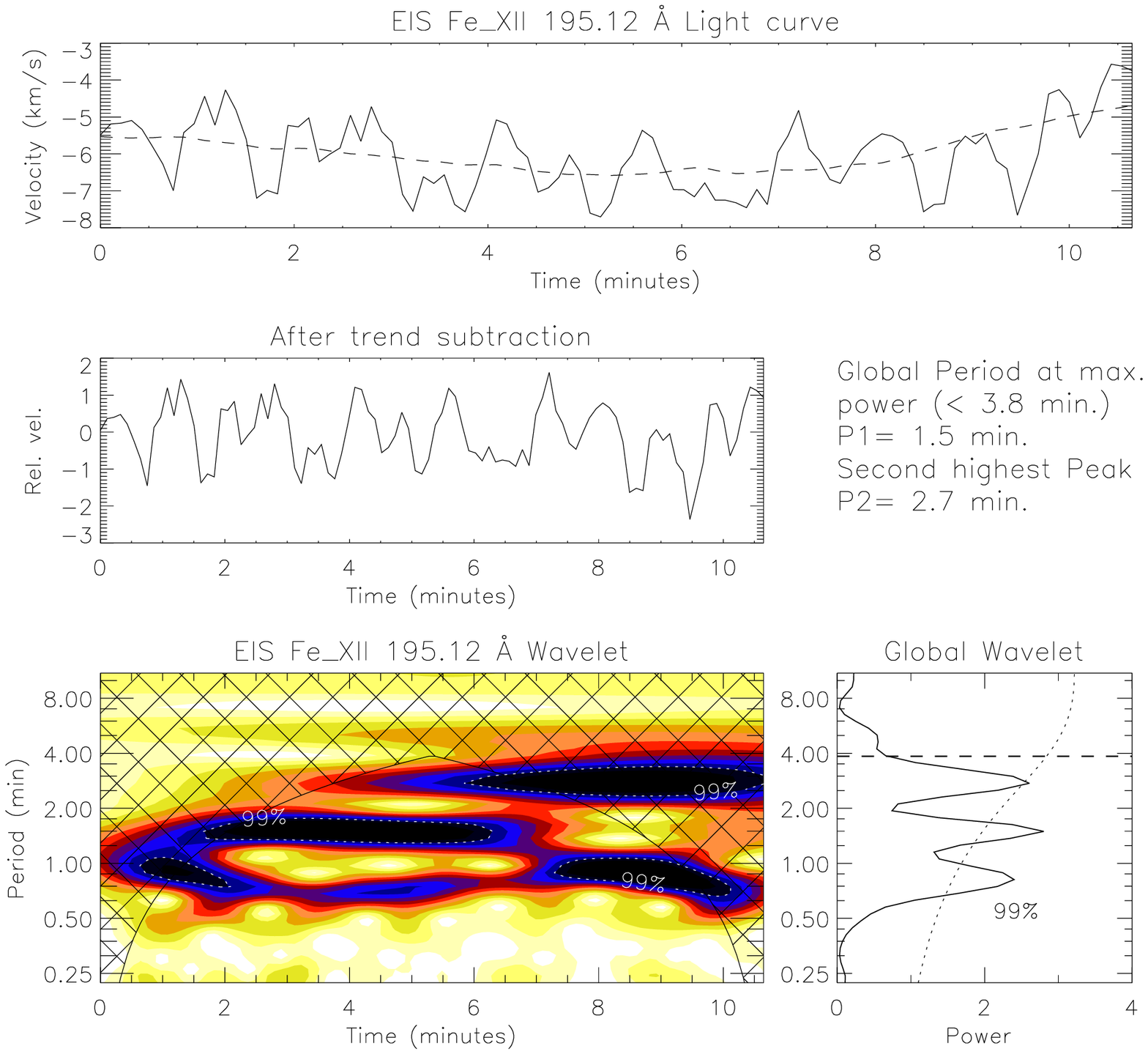}\\
 \vspace*{0.1in}
 \includegraphics[width=9.0cm,angle=90]{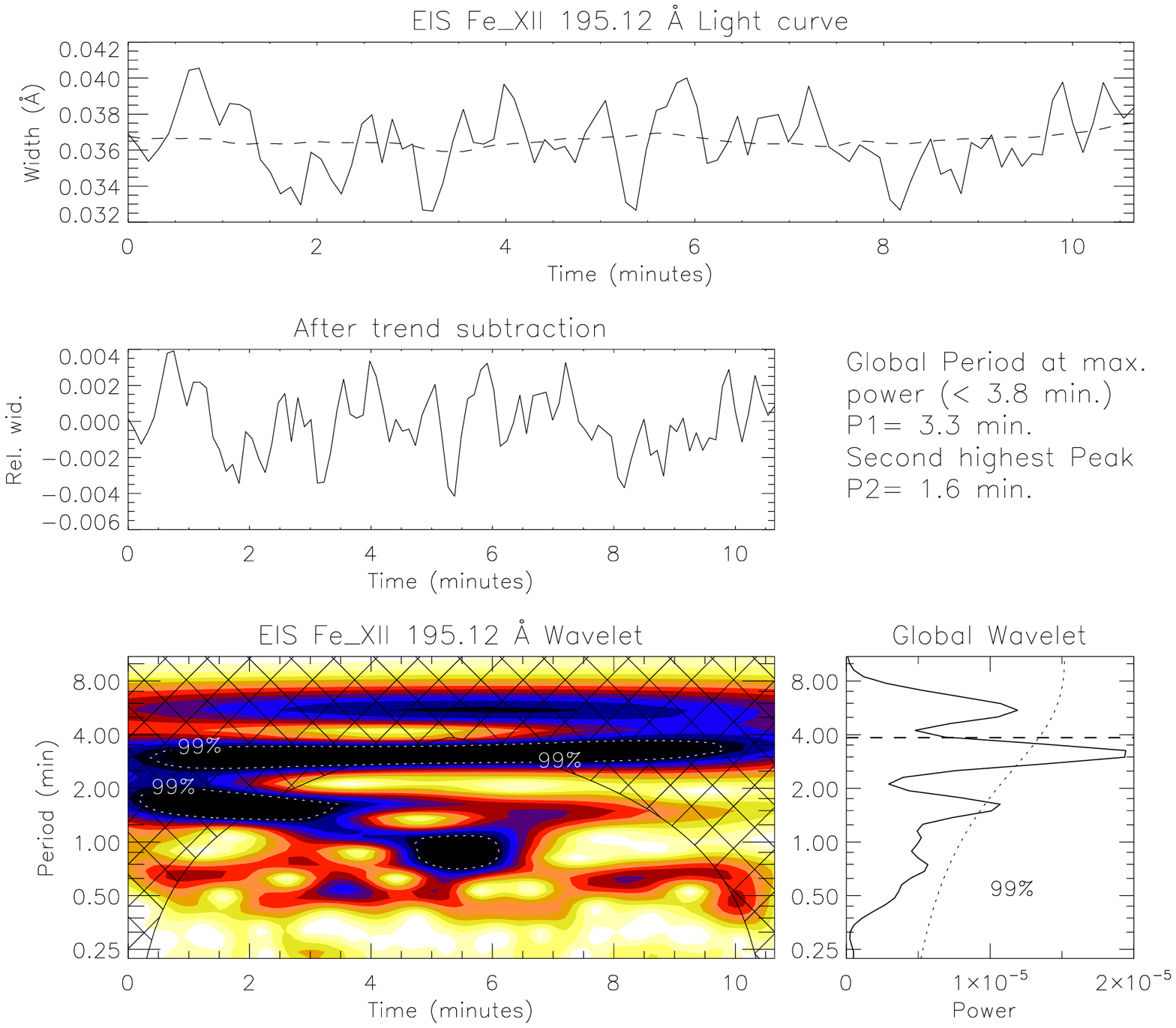}
\caption{Continued.}
\label{waveplots1}
\end{figure}

We use the wavelet technique (\opencite{1998BAMS...79...61T}) to find the periodicities present in these light curves.  Figure~\ref{waveplots1} shows some of the results. These plots show results for intensity, Doppler shift and line width in EIS \FeXII\ 195.12~\AA\ line and intensity in AIA 193~\AA\ channel (See the corresponding labels to identify individual plots). Each light curve from EIS in intensity, Doppler shift, and line width, was smoothed over two temporal points before proceeding for wavelet analysis. This removes any variations shorter than $\approx$12~s.  For AIA, the light curves are directly used without doing any such smoothing. These are shown in the top panels of each plot in Figure~\ref{waveplots1}. All of them clearly show some oscillatory behavior. A 40 point ($\approx$ 4~min) running average for EIS and 20 point ($\approx$ 4~min) for AIA light curves, is subtracted from these to filter the longer periods. In all the cases discussed in this article the long period filtering is chosen such that we do not eliminate any reliable periods from EIS and the equivalent filtering scale is followed in AIA. The subtracted trend is shown with a dashed line overplotted on the light curves in the top panels. The resultant light curves are displayed in the middle panel of each plot. For the intensity, the resultant light curves are also normalized by the subtracted trend and displayed here. These are input to the wavelet program for identification of various periods. The wavelet results are displayed in the bottom left panels of each plot in this figure. This shows the different periodicities present and their temporal evolution. The cross hatched region in this plot is the region that can suffer from edge effects and hence the periods observed in this region are not reliable. This region is called Cone of Influence (CoI; \opencite{1998BAMS...79...61T}). The power at different periods is averaged over time and displayed in  the bottom right panels of each plot. This is called the global wavelet plot which shows the strength of different frequencies for the total duration. The dashed horizontal line in this plot marks the cut off for the longest period that can be observed from this data. Different peaks observed in this plot indicate simultaneous presence of different periodicities. The confidence of detection of these periods is measured by calculating a significance level assuming a white noise process (\opencite{1998BAMS...79...61T}). A curved dotted line overplotted on this plot marks this level for 99\% significance. This essentially means that the probability that any peak above this level is real, is 99\%. From this plot we then identify the peak periodicity values above this level. The top two strongest periodicities are described right to the central panels. We find an oscillation with periodicity roughly around 1.6~min (see Figure~\ref{waveplots1}) in all the three line parameters intensity, velocity, and line width with 99\% confidence. Analysis using other EIS line \FeXIII\ 202.04~\AA\ shows similar periodicity peaking at 1.2~min in all the three line parameters. Corresponding results from AIA 193~\AA\ channel show two periods peaking at 3.9~min and 6.1~min. The period at 6.1~min has a larger power and hence was not eliminated completely, though its power reduces due to 4~min filtering. We also find a period at 1.7~min, but it is not significant. In 171~\AA\ channel we observe two peaks at 4.7~min and 2.0~min, of which, again the latter is below the confidence level.
\subsection{Long period oscillations}
 \label{S-Long period oscillations}
The location marked by the top horizontal cut in Figure~\ref{marks} (solar-$Y \approx$ -226\arcsec), shows oscillations of longer periods, roughly of the order of 9~min. This pixel location falls over the portion of fan loops with no visible background structure. Since, the duration of each sit-and-stare set is around 10~min, we combined all the three sets to observe such long periods. The short time gap (7~s to 8~s) between each set allowed us to do this directly. There are 8 missing frames, not all consecutive, probably due to telemetry loss, in the first set which are linearly interpolated. Light curves from EIS in intensity, velocity, and line width were constructed from this pixel location and were smoothed over 8 temporal points before applying wavelet analysis to find the periodicities. For AIA, we take a square region similar to that explained in Section~\ref{S-Short period oscillations} to generate the light curves. A running average constructed by smoothing over 110 points ($\approx$11~min) in EIS and equivalent in AIA (55 points) is subtracted from these light curves to filter periodicities longer than 11~min. Wavelet analysis results in EIS 195.12~\AA\ line and AIA 193~\AA\ channel are shown in Figure~\ref{waveplots2}.
\begin{figure}
 \centering
 \includegraphics[width=9.0cm,angle=90]{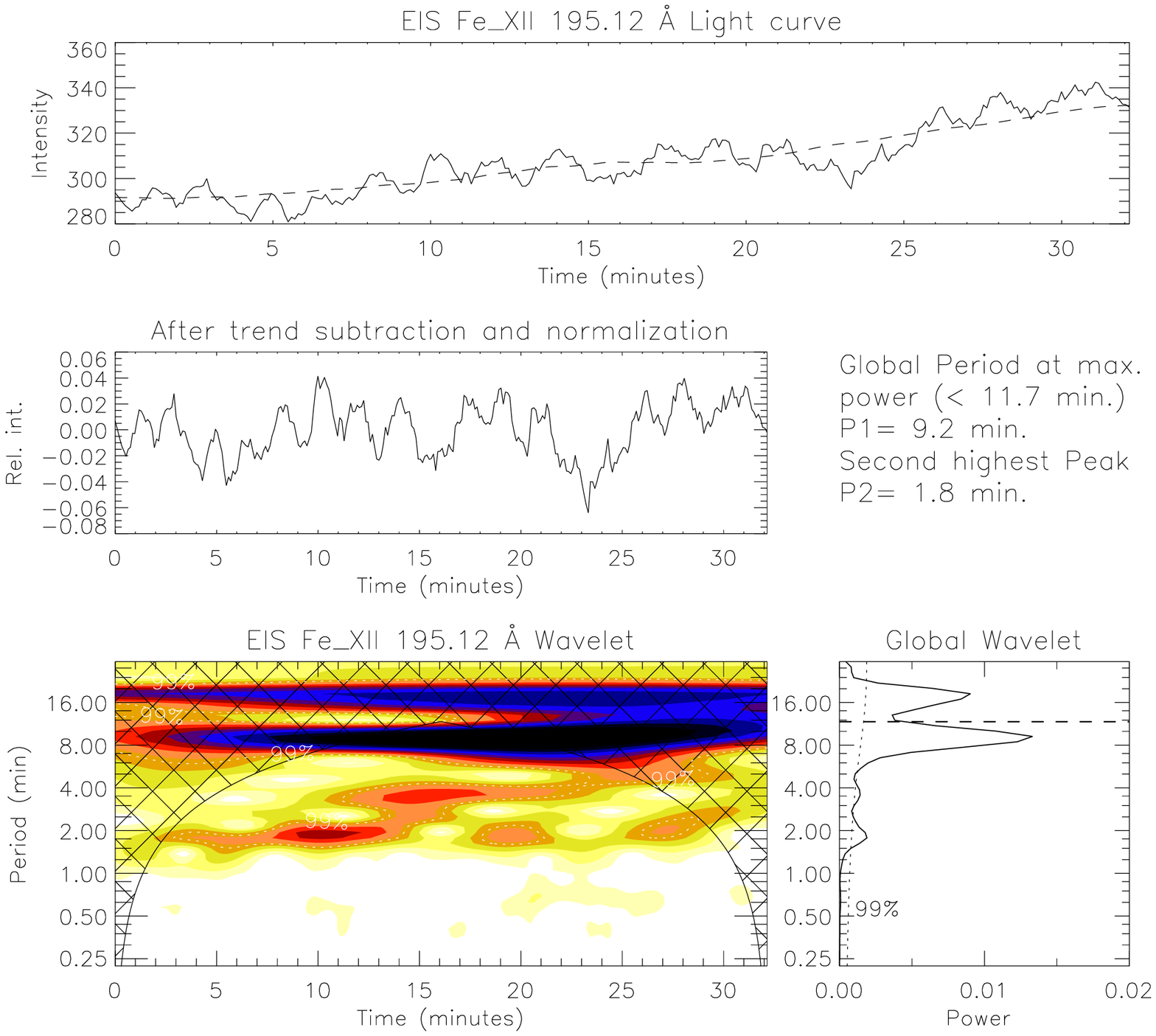}\\
 \vspace*{0.1in}
 \includegraphics[width=9.0cm,angle=90]{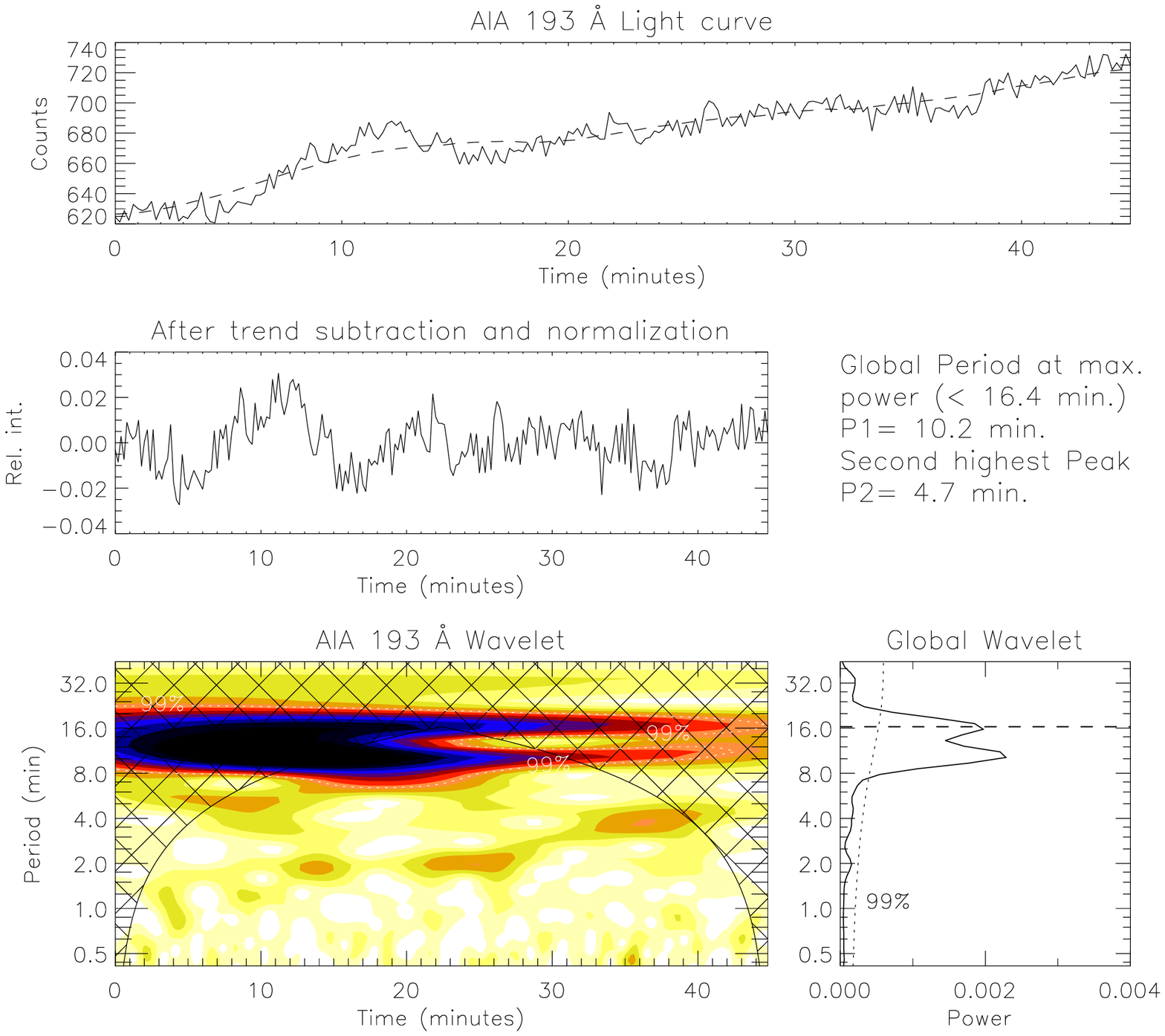}
 \caption{Wavelet analysis results for the location marked by the top cut (solar-$Y \approx$ -226\arcsec) in Figure~\ref{marks}. These plots are for intensity, velocity and line width in \FeXII\ 195.12~\AA\ EIS line and intensity in 193~\AA\ channel of AIA (see the corresponding labels). Original light curves shown in the top panels are smoothed over 8 points for EIS and no smoothing is done for AIA. The dashed lines overplotted on these are made by taking a 110 point ($\approx$11~min) running average of the original for EIS and 55 point (11~min) for AIA. See Figure~\ref{waveplots1} for the description of individual panels.}
\end{figure}
\begin{figure}
\ContinuedFloat
 \centering
 \includegraphics[width=9.0cm,angle=90]{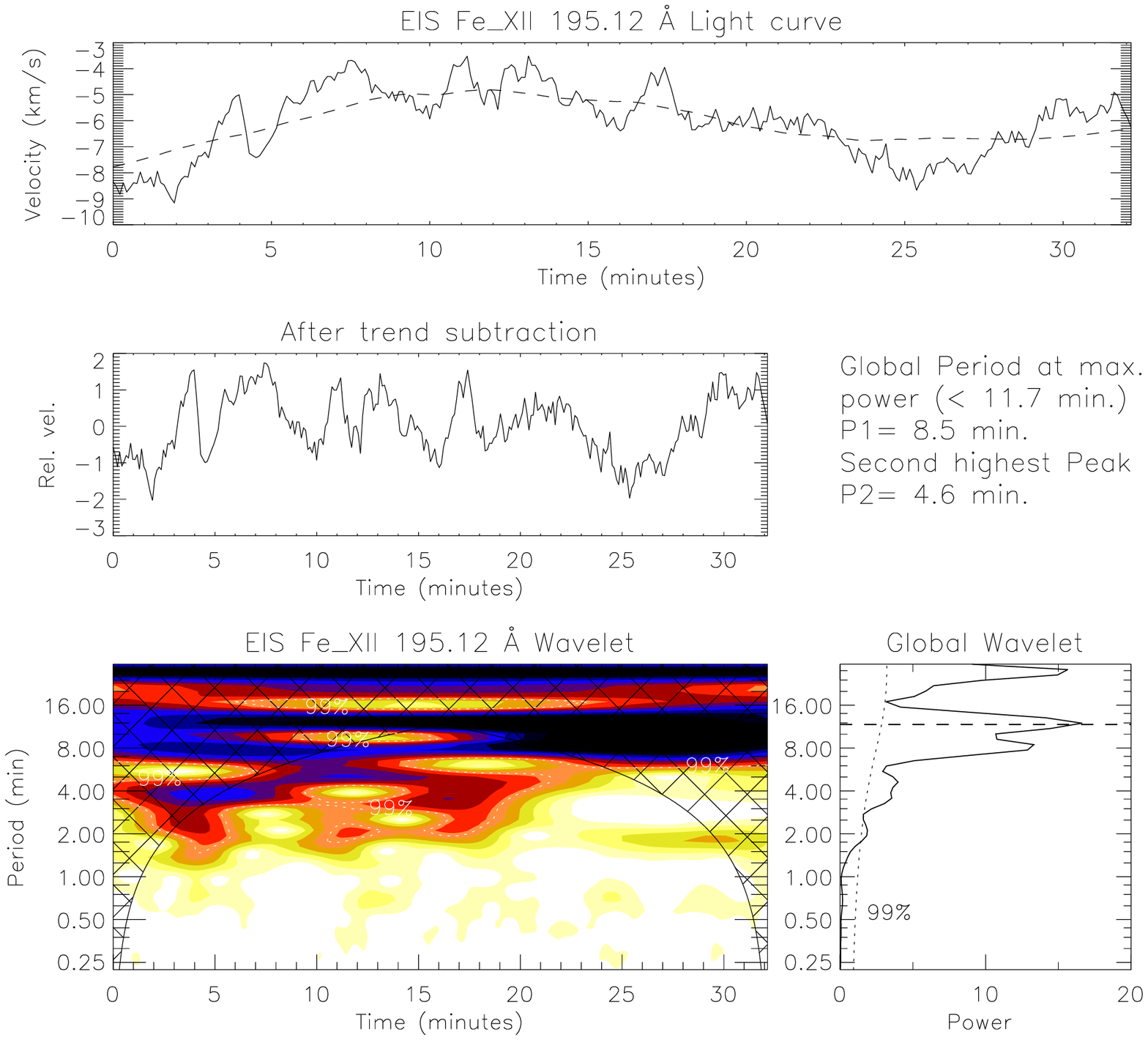}\\
 \vspace*{0.1in}
 \includegraphics[width=9.0cm,angle=90]{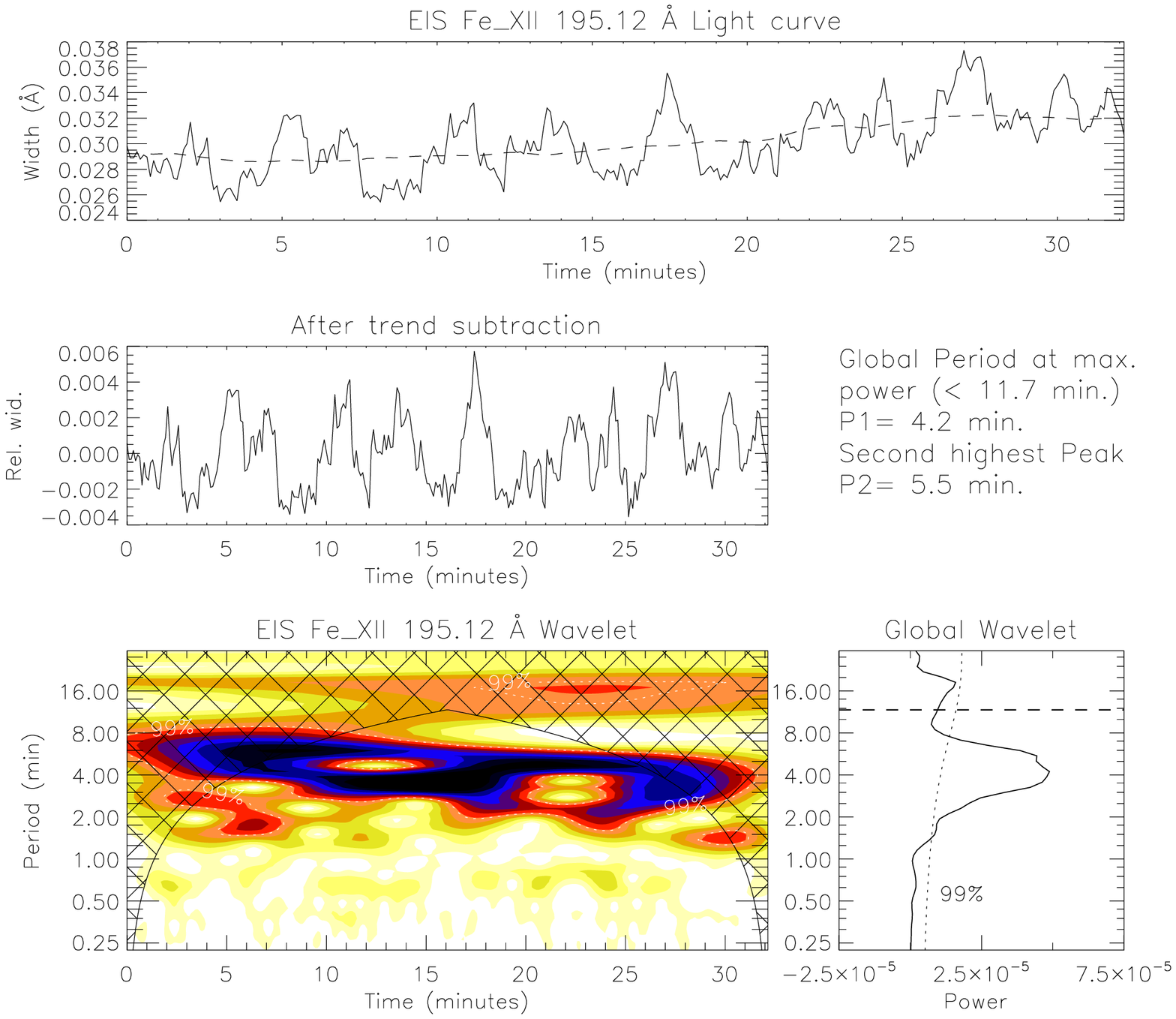}
\caption{Continued.}
\label{waveplots2}
\end{figure}
Global wavelet plots show a period around 9~min with 99\% confidence both in intensity and velocity, but not in line width. It may be noted that the plot in intensity from AIA 193~\AA\ channel shows a broader peak at 10.2~min which may or may not correspond to this period. However, results from 171~\AA\ channel show this period clearly peaking at 9.4~min. Though not significant, a period at 4.7~min is observed in both channels.

\begin{figure}
 \centering
 \includegraphics[width=9.0cm,angle=90]{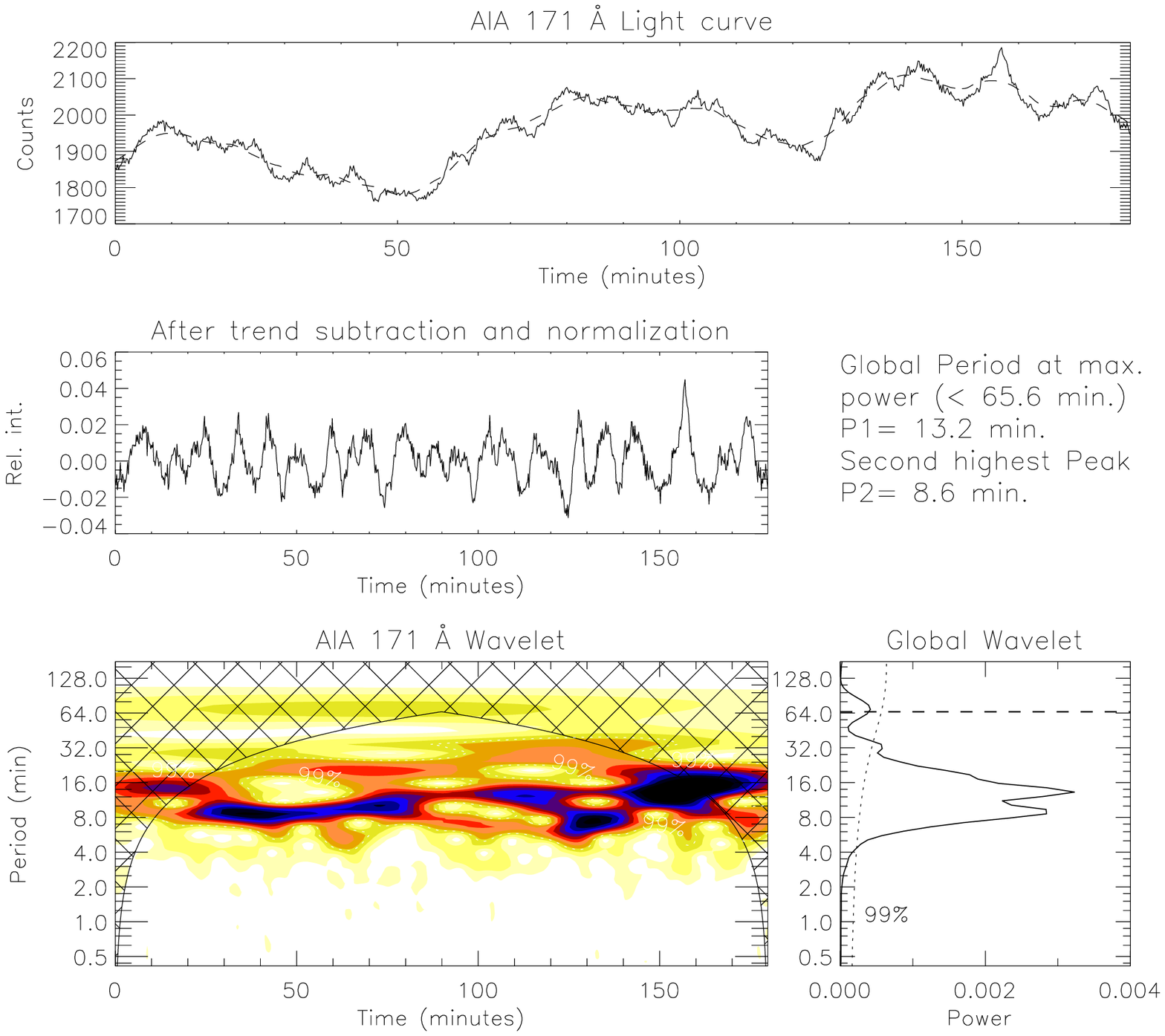}\\
 \vspace*{0.2in}
 \includegraphics[width=9.0cm,angle=90]{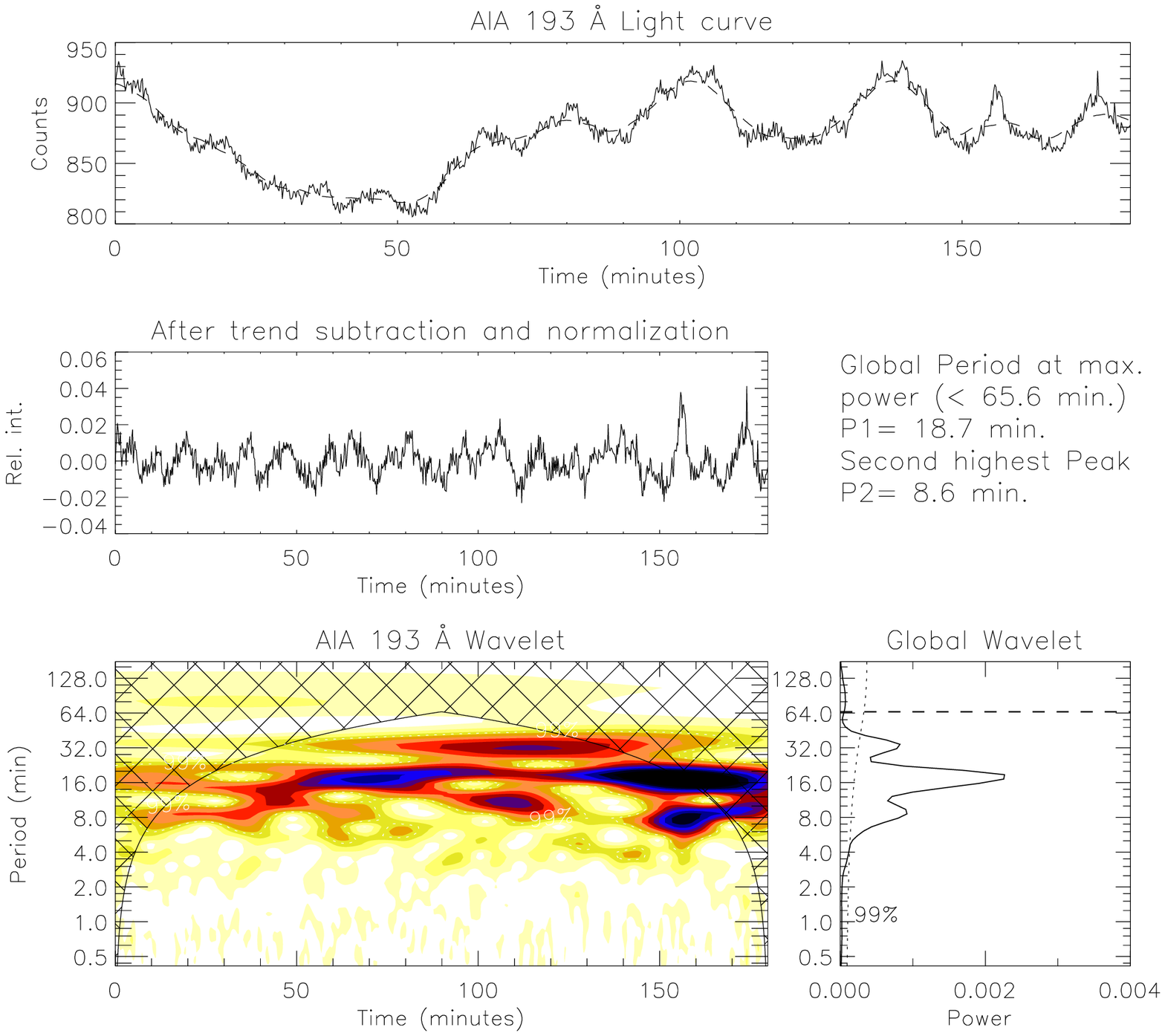}
\caption{Wavelet results for the light curves from 171~\AA\ (top) and 193~\AA\ (bottom) channels of AIA. These are generated from an arbitrarily chosen pixel location which is 30\arcsec\ far from the foot point of the loop bounded by the dashed curves shown in Figure~\ref{marks}. The light curves in the middle panels are made by detrending and normalizing the originals (top panels) with a 55 point (11~min) running average. Overplotted dashed line in each top panel shows this running average. Description of individual panels is same as that in Figure~\ref{waveplots1}.}
\label{waveplots3}
\end{figure}

To verify this further and extract more information, we use the imaging sequence from AIA in 171~\AA\ and 193~\AA\ channels. The duration of this dataset is 3~hrs covering the EIS observation time. More details on preparation of this data is given in Section~\ref{S-Observations}. We follow a portion of the fan loop system that crosses the slit and encloses the pixel location marked by the top cut (solar-$Y \approx$ -226\arcsec) in Figure~\ref{marks}. Dashed curves in this figure bound the chosen loop portion. We follow an analysis similar to that described in \inlinecite{2000A&A...355L..23D}, to construct a space-time map for this region. We divide this region into several cross-sections depending on the average length of the loop. Counts in each cross-section are summed and normalized to the number of pixels in that region. This gives us a 1-d array of average counts along the length of the loop. All such arrays from different snapshots are stacked together to construct a space-time map. We then process these space-time maps by detrending and normalizing, to enhance the fine variations. We use 55 point smoothing to detrend which should normally filter out all the periods longer than 11~min. 
\begin{figure}
 \includegraphics[width=3.4cm,angle=90]{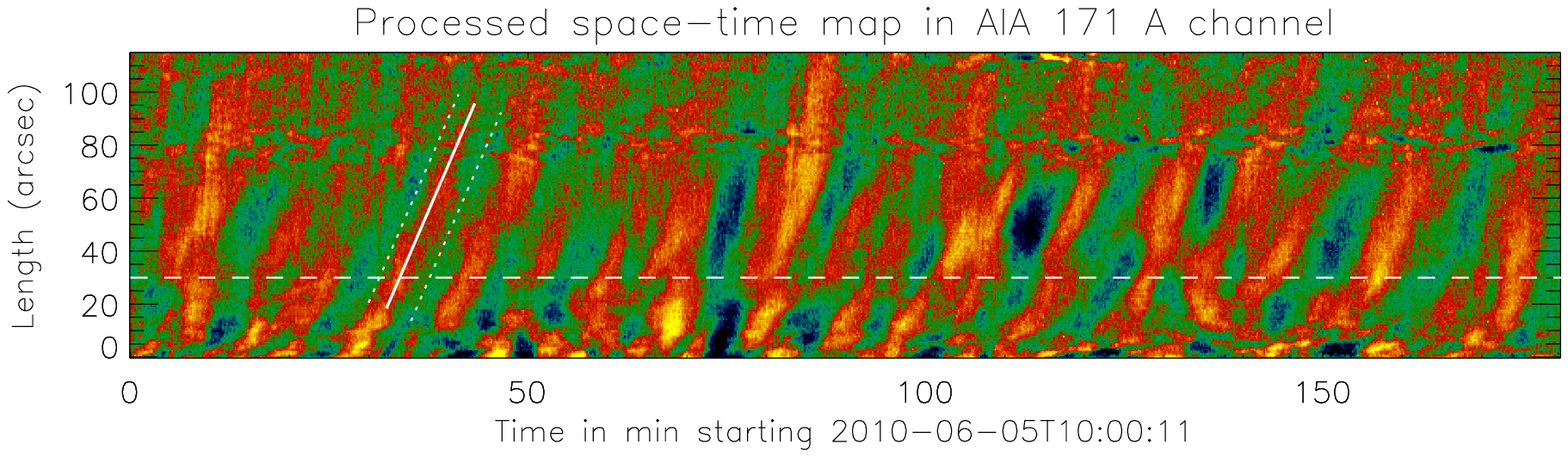}\\
 \includegraphics[width=3.4cm,angle=90]{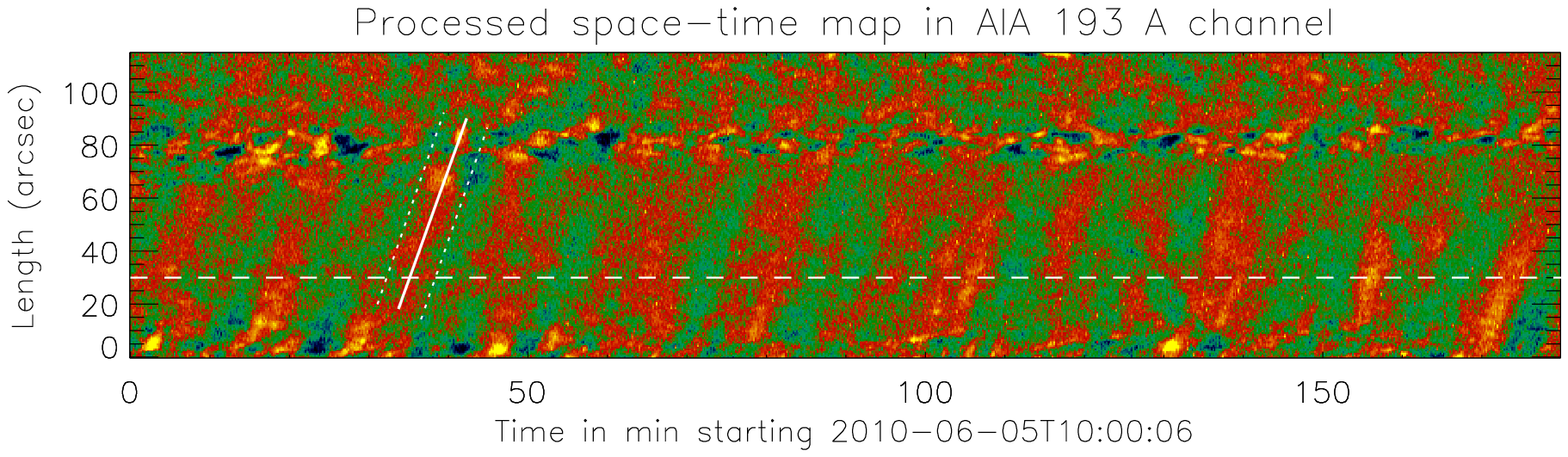}
\caption{Processed space-time maps in 171~\AA\ (top) and 193~\AA\ (bottom) channels of AIA constructed from the loop region bounded by dashed curves in Figure~\ref{marks}. Alternating ridges with positive slope indicate a quasi-periodic disturbance propagating outward. The dashed horizontal line marks the center of 3 adjacent rows used in the wavelet analysis for periodicity estimation. The slanted dotted lines in each of these plots, bound the strip considered to identify peaks. The solid line is the linear fit of all the peak positions along the strip. The slope of this gives the propagation speed value. }
\label{speed}
\end{figure}
Figure~\ref{speed} shows these processed space-time maps in 171~\AA\ and 193~\AA\ channels of AIA. Alternating slanted ridges, representing the propagating disturbances, are visible in both the channels. Positive slope indicates that they are propagating outward. We also did this analysis in the hotter 211~\AA\ channel of AIA but these disturbances are hardly visible. To measure the periodicity we constructed a light curve from the original space-time map by averaging over three rows centered at the row marked by the horizontal dashed line in Figure~\ref{speed}. This is 30\arcsec\ above the foot point of the loop and is arbitrarily chosen. Wavelet analysis results for these light curves in both AIA channels are shown in Figure~\ref{waveplots3}.
We find a period peaking at 8.6~min in both channels which is close to the $\approx$9~min period we observed at EIS pixel location showing longer periods. Also the disturbances in the space-time maps can be clearly seen up to the end of the chosen loop portion. So, it is more likely that these are one and the same. We can also measure the apparent propagation speed of these disturbances from the slope of slanted ridges in the space-time maps. This is done by following the method described in the next subsection.
\subsubsection{Propagation speed}
Any periodic propagating disturbance will appear as alternating ridges of brightness with finite slope, in a space-time map. The amplitude of the disturbance is normally low and it requires some processing to remove the background and enhance the visibility of these ridges. The slope of these ridges gives the apparent propagation speed. This is one of the important observational parameters that help to understand the nature of the detected oscillations. For instance, the temperature dependence of the speed indicates the acoustic nature of the observed oscillations. But to establish this, we need considerably precise measurements of the speed in at least two different channels representing different temperature plasma. The traditional way to calculate the speed is to manually select two points on one of the clean ridges, connect them with a straight line and find the slope of that line. This roughly gives the propagation speed, but the error in this estimation is considerably high. Here we attempt to reduce 
this error, by following a more rigorous method. First we extract a clean ridge from the space-time map, (processed to display the alternate ridges) by choosing a strip roughly parallel to it and sufficiently wide to only include this particular ridge. The dashed lines in each panel of the Figure~\ref{speed} enclose such strips. Then we find the positions of the local maxima at each spatial location along the strip. These positions are then converted back to the actual positions in the space-time map using co-ordinate transformations. Ideally, for a clean oscillatory disturbance propagating with constant speed, these should fall on a straight line and the slope of that line gives the propagation speed with no error. But the presence of some background structures, lower amplitudes, and/or multiple periodicities can affect these positions and deviate them significantly. Nevertheless, most of them fall on a straight line. 
\begin{figure}
 \centering
 \includegraphics[width=11cm]{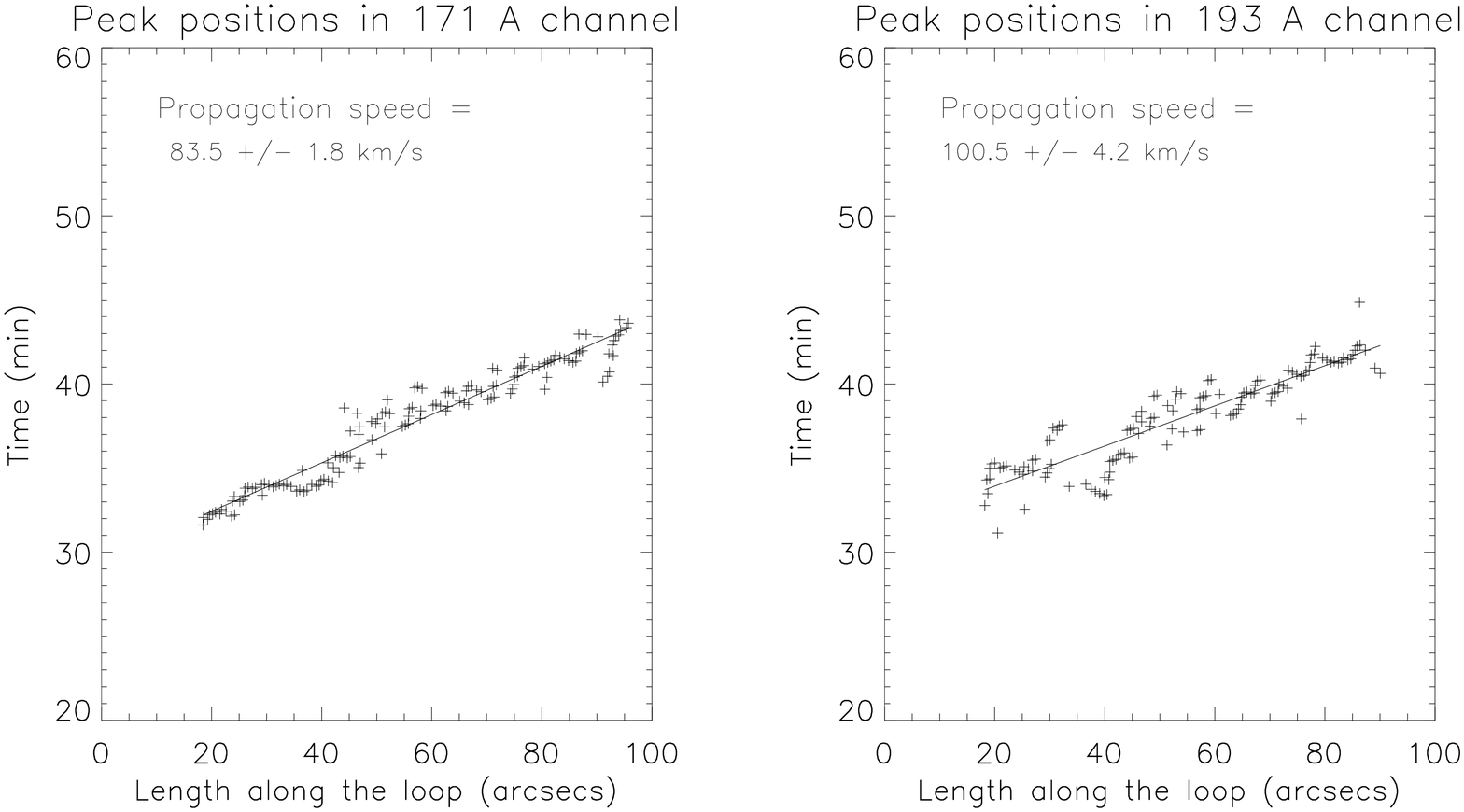}
\caption{Results of the propagation speed analysis in AIA 171~\AA\ (left) and 193~\AA\ (right) channels. In each plot '+' marks denote the observed positions of the peaks along the strip bounded by dotted lines in Figure~\ref{speed}. The overplotted solid line is the linear fit. The inverse slope of this line gives the propagation speed. Estimated propagation speeds are discussed in the text.}
\label{peakpos}
\end{figure}
Figure~\ref{peakpos} displays the observed positions of the peaks along the chosen strips in 171~\AA\ (left) and 193~\AA\ (right) channels which are denoted by '+' symbols. The overplotted solid line in each plot, is the linear fit. The positions derived from this linear fit are marked by a continuous line on the chosen strip in Figure~\ref{speed}. The slope of this line gives the propagation speed. The propagation speeds thus measured are 83.5 $\pm$ 1.8~\kms\ in 171~\AA\ and 100.5 $\pm$ 4.2~\kms in 193~\AA\ channels. Error values in these speed estimations are calculated from the 1-$\sigma$ error in the linear fit. Propagation speeds calculated with this method are more reliable with considerably small errors. Since we identify the peaks from the local maxima, this method requires a clean continuous ridge with good contrast between the peak values and the background, in other words amplitudes should be sufficiently large. \\

\inlinecite{2009ApJ...697.1384T} used a similar method to calculate the propagation speeds using the time lag at different positions along the ridge obtained from the cross-correlation between the time series. The accuracy in their measurement is also similar to that obtained in the current method. The correlation may not be good all the time if we consider the full time series and the additional advantage in our method by isolating individual ridges is that any change in the speed in the successive disturbances because of some additional events can also be explored and compared.

\section{Discussion} 
 \label{S-Discussion}
In this section, we discuss the possible interpretation for the observed periodicities. Table~\ref{prds} summarizes the different periods observed both from EIS/{\it Hinode} and AIA/SDO at the top (solar-$Y \approx$ -226\arcsec) and bottom cut (solar-$Y \approx$ -250\arcsec) locations in Figure~\ref{marks}. Here we list the top two strongest periods with 99\% confidence from the wavelet analysis. Values that are not significant are marked with (*).

\begin{table}
\caption{Top two strongest periods above the 99\% significance level obtained from wavelet analysis. Results from the two AIA channels and the two EIS lines at both locations marked by horizontal cuts in Figure~\ref{marks} are listed. Values in brackets correspond to the secondary peaks.}
 \begin{tabular}{c c c c c}     
 \hline
   Solar-$Y$  & Emission line/       & Intensity & Velocity & Line width \\
              &  Channel             &           &          &  \\
 \hline
  -226\arcsec & \FeXII\ 195.12~\AA\  & 9.2  (1.8)  & 8.5 (4.6) & 4.2 (5.5) \\
              & \FeXIII\ 202.04~\AA\ & 7.1  (2.1)  & 6.0 (3.3) & 3.9 (10.1) \\   
              & AIA 171~\AA\         & 9.4  (4.7*) & ---       & ---      \\
              & AIA 193~\AA\         & 10.2 (4.7*) & ---       & ---      \\
  -250\arcsec & \FeXII\ 195.12~\AA\  & 1.6  (0.82) & 1.5 (2.7) & 3.3 (1.6) \\
              & \FeXIII\ 202.04~\AA\ & 1.2  (0.82) & 1.4 (2.3) & 2.7 (1.2) \\   
              & AIA 171~\AA\         & 4.7  (2.0*) & ---       & ---      \\
              & AIA 193~\AA\         & 3.9  (6.1) & ---       & ---      \\
 \hline
 \end{tabular}
\label{prds}
\newline
{\bf Note:} Values marked by (*) are not significant.
\end{table}

At the pixel location corresponding to solar-$Y \approx$ -250\arcsec~, we find short period oscillations. Results from EIS in \FeXII\ 195.12~\AA\ and \FeXIII\ 202.04~\AA\ lines show a periodicity peaking at $\approx$ 1.6~min and $\approx$ 1.2~min, respectively, in all the three line parameters. Similar periodicities are also seen from AIA, but they are not significant in either of the channels. This might be due to the wider passbands of AIA filters (compared to the narrow spectral lines of EIS) covering a larger temperature range, which allows more background in the line of sight and may result in reduction of the relative oscillation amplitudes. This can cause problem in detecting, particularly, short period oscillations whose amplitudes are normally less. Presence of significant oscillations in all the three line parameters, intensity, velocity and line width makes it difficult to point a particular wave mode as the cause. \inlinecite{2010ApJ...722.1013D} discussed that a faint blue-shifted emission 
component in the line profile, caused by quasi-periodic high velocity upflows, can actually cause these small amplitude oscillations, in intensity, velocity and line width when fitted with single Gaussian. Moreover, this particular pixel location is adjacent to a bright structure which is close to the active region core and could possibly be the foot-point of a hotter loop where the chances of upflows are more probable. In fact, there was a small loop system at this location, 1 hour earlier to this observation, which disappeared later leaving this bright structure. To investigate this possibility, we browsed through several line profiles at this location but found no visible blue shifted component. We also constructed a mean line profile taking the average over time which is shown in Figure~\ref{profiles} (bottom row) for both EIS lines. Since our analysis at this location is only on the third set in sit-and-stare mode we restrict our time average to this. The solid lines are the single Gaussian fits to the profiles considering a polynomial background. Presence of any additional periodic component should be visible in this profile, although averaging can reduce its amplitude. It can be clearly seen from this figure that there is no enhancement in the blue wing of the profile and a single Gaussian fit is a very good approximation. So, the possible explanation of these oscillations in terms of quasi-periodic upflows may be ruled out. However, there can be continuous upflows in the background. Doppler shift values (see Figure~\ref{waveplots1}) indicate that the line of sight speeds of such flows at this location are less than 10~\kms. \inlinecite{2011SoPh..270..213S} also found oscillations of very short periods (25~s to 50~s), from eclipse observations, in all the three line parameters. They discussed this in terms of possible coupling of more than one wave mode. Also in our observation, we find a period $\approx$3~min only in velocity and line width and another period $\approx$0.8~min significant only in intensity and velocity. So, it is quite possible that these oscillations are due to coupling of various MHD modes at this location.

\paragraph*{}
We find long period oscillations at the location corresponding to solar-$Y \approx$ -226\arcsec. Results from EIS \FeXII\ 195.12 line show a periodicity around 9~min in both intensity and velocity, but not in line width. Intensities from both 171~\AA\ and 193~\AA\ channels of AIA, also show this periodicity. EIS \FeXIII\ 202.04 line shows different peaks ranging from 6~min to 10~min (refer to Table~\ref{prds}) in intensity, velocity and line width which may or may not be related to this $\approx$9~min period. Space-time analysis over a loop crossing this region shows propagating disturbances with periodicity 8.6~min in both the AIA channels. Mean line profiles at this location are shown in Figure~\ref{profiles} (top row) in both the EIS lines. These profiles are constructed by averaging over all the three sit-and-stare sets to compare with these results. Single Gaussian fits, along with a polynomial background, to these profiles are overplotted as solid lines. There is no apparent blue shifted component (deviation from the single Gaussian fit) visible from these profiles. Oscillations only in intensity and velocity with no corresponding peak in line width and no visible blue-shifted emission in the line profiles, suggest that the observed long period oscillations can be of magneto-acoustic type. Also the apparent propagation speeds are calculated as 83.5$\pm$1.8~\kms\ and 100.5$\pm$4.2~\kms\ respectively, in 171~\AA\ and 193~\AA\ channels. These are significantly lower than the theoretical acoustic speeds, 136~\kms\ and 170~\kms\, taking peak temperatures 0.8~MK and 1.25~MK respectively, in these channels. Also, the ratio of the observed speeds is 1.20 compared to the theoretical value 1.25. This may suggest that these are propagating slow magneto-acoustic oscillations. However, we do not see any clear correlation between intensity and Doppler shift which is expected in case of propagating slow waves. Although they go in phase at a few locations, overall the correlation seems to be poor. It is possible that the presence of higher frequencies, which do not have correlated variations in Doppler shift, is spoiling the overall correlation. It is also important to note that the propagation speeds in both channels are calculated from the same but single disturbance. It may better to compare the speeds for more than one propagating disturbance, but it is very hard to find a clean continuous ridge of sufficient length, in this dataset, simultaneously in both channels. Particularly in the 193~\AA\ channel the amplitudes were small and the ridges were mostly bad making our attempts futile.

\paragraph*{}
\inlinecite{2009ApJ...696.1448W} studied oscillations in different temperature lines and found that the oscillation amplitudes decrease with increase of temperature. They explained this in terms of damping due to thermal conduction. \inlinecite{2010ApJ...713..573M} showed several cases of damped oscillations in different temperature lines. They observed a mixed behavior, that in some cases, the oscillation amplitude increases with increasing line formation temperature and in other cases it decreases. We observe undamped oscillations over several cycles and find that the relative amplitude of the oscillations, in EIS \FeXIII\ 202.04 line is higher than that in \FeXII\ 195.12 line in all the line parameters, at the location corresponding to solar-$Y \approx$ -250\arcsec. It is more difficult to explain this behavior, since the oscillations at this location do not correspond to a particular mode. The oscillation amplitudes at the other location (solar-$Y \approx$ -226\arcsec) showing long period oscillations cannot be compared directly, as the periodicities are different in different EIS lines (see Table~\ref{prds}). However, assuming that these oscillations are caused by the propagating disturbances observed in fan loops we can compare the amplitudes in different channels of AIA (see Figure~\ref{waveplots3}). It can be seen that the amplitudes are lower in the hotter 193~\AA\ channel (see Figure~\ref{waveplots3}). This is consistent with the damping in a multi-thermal loop due to thermal conduction and may further support the nature of these oscillations as of slow magneto-acoustic type.

Figure~\ref{waveplots3} shows a period at $\approx$18~min along with the $\approx$9~min period, in both channels. It is not so clear in the 171~\AA\ channel because of the 11~min filtering, but it can be clearly seen by filtering periods longer than 20~min. This may suggest the presence of harmonics in the propagating disturbances, similar to those observed by \inlinecite{2009A&A...503L..25W} who report 12~min and 25~min periods in the active region fan loops.

\section{Conclusions} 
We searched for oscillations in active region fan loops using simultaneous high cadence spectroscopic (EIS/{\it Hinode}) and imaging (AIA/SDO) data. We find two locations showing oscillations of short ($<$1~min to 3~min) and long periods ($\approx$9~min). Short periods are observed with oscillations in all the three line parameters, intensity, Doppler shift, and line width which is possibly due to coupling of different MHD modes. Amplitudes of these oscillations, in all the line parameters, show significantly higher values in higher temperature line. Investigation at the same location from AIA does not show these periods with significant peaks. The amplitudes of these periods from AIA are very low and it is difficult to observe reliable oscillations with periods less than 2~min from AIA. This might be due to the increased background emission in the line of sight from the wider passbands of AIA. Longer periods identified over the fan loops show oscillations only in intensity and Doppler shift but not in line width. Space-time analysis from AIA over a loop region covering this pixel location, indicates oscillations with similar periods propagating with different speeds in different temperature channels. Recently, \inlinecite{2012SoPh..279..427K} also find that velocities of the propagating disturbances in structures located at sunspot regions are temperature dependent. The apparent propagation speeds are 83.5$\pm$1.8~\kms\ and 100.5$\pm$4.2~\kms\ in 171~\AA\ and 193~\AA\ channels, respectively. The ratio of these speeds, is close to the theoretical acoustic speed ratio in this region. This might indicate that these oscillations are due to slow magneto-acoustic waves. These waves also show lower amplitudes in the hotter channel of AIA consistent with damping due to thermal conduction. Line profiles at both these locations do not show any visible blue-shifted emission component. The asymmetry in the line profiles due to any possible periodic upflows can be masked if the field lines are highly inclined to the line of sight \cite{2012ApJ...748..106T}. But the observation of significant Doppler shift oscillations at these locations might indicate that this is not the case here. We require detailed spectroscopy to completely resolve the ambiguity and the forthcoming IRIS mission can be very useful to address this.

\begin{acks}
We would like to thank the referees for their useful comments. Hinode is a Japanese mission developed and launched by ISAS/JAXA, with NAOJ as domestic partner and NASA and STFC (UK) as international partners. It is operated by these agencies in co-operation with ESA and NSC (Norway). The AIA data used here is the courtesy of SDO (NASA) and AIA consortium.
\end{acks}

%%% BIBLIOGRAPHY %%%%%%%%%%%%%%%%%%%%%%%%%%%%%%%%%%%%%%%%%%%%%%%%%%%%%%%%%%%
     % format of references provided by the journal (.bst)
\bibliographystyle{spr-mp-sola}
\bibliography{kpref.bib}  

\end{article} 

\end{document}